\documentclass[aps,pre,reprint,twocolumn,superscriptaddress,showpacs,floatfix]{revtex4-1}
\usepackage[latin1]{inputenc}
\usepackage{epsfig}
\usepackage{amsfonts}
\usepackage{amssymb}
\usepackage{calrsfs}
\usepackage{amsmath}
\usepackage{xcolor}
\usepackage{graphicx}
\usepackage[colorlinks=true,allcolors=blue]{hyperref}
\hypersetup{breaklinks=true}

\begin{document}
	
	\title{Oscillating states of periodically driven anharmonic Langevin systems}
	\author{Shakul Awasthi}
	\email{shakulawasthi010615@iisertvm.ac.in}
	\author{Sreedhar B. Dutta}
	\email{sbdutta@iisertvm.ac.in}
	\affiliation{School of Physics, Indian Institute of Science Education and Research Thiruvananthapuram, India}
	
	\date{\today}
	
	\begin{abstract} 
		{We investigate the asymptotic distributions of periodically driven anharmonic Langevin systems. Utilizing the underlying $SL_2$~symmetry of the Langevin dynamics, we develop a perturbative scheme in which the effect of periodic driving can be treated nonperturbatively to any order of perturbation in anharmonicity. We spell out the conditions under which the asymptotic distributions exist and are periodic, and show that the distributions can be determined exactly in terms of the solutions of the associated Hill equations. We further find that the oscillating states of these driven systems are stable against anharmonic perturbations.}
	\end{abstract}

	\maketitle
	
	\newcommand{\bee}{\begin{equation}}
		\newcommand{\eee}{\end{equation}}
	\newcommand{\tm}{(t)}
	\newcommand{\gm}{\gamma}
	\newcommand{\dbar}{d\hspace*{-0.08em}\bar{}\hspace*{0.1em}}
	\newcommand{\haf}{\frac{1}{2}}
	\newcommand{\x}{\mathbf{x}}
	
	%
	%
	%
	%
	
	\section{Introduction}\label{intro}
	
	The study of the effects of periodic forces on a variety of systems, including classical~\cite{Higashikawa2018,Salerno2016}, quantum~\cite{Bukov2015,Eckardt2015,Kohler1997,Lewis1969,Brandner2016} and statistical systems~\cite{Jung1993,Brandner2015,Dutta2003,Dutta2004,Wang2015,Knoch2019,Gammaitoni1998,Kim2010,Fiore2019,Koyuk2018,Oberreiter2019,Tociu2019}, has been of continual interest for various and diverse reasons.
	
	Periodically driven many-particle systems, under certain conditions, can exist in a state exhibiting periodic thermodynamic properties. It may not be possible to uncover the features of this {\it oscillating state} either from the knowledge of the equilibrium properties of the corresponding systems in the absence of driving or by studying the effects of weak periodic forces on the thermodynamics of such systems. Instead it may be required to explore the behavior  of such systems by necessarily treating the driving nonperturbatively.

	A suitable framework to investigate the properties of an oscillating state is presumably some sort of stochastic thermodynamics, wherein the periodic driving is appropriately incorporated. The fluctuations of a macroscopic variable of a thermodynamic system can be assumed to follow a continuous Markov process even in the presence of driving. It is expected that the stochastic process that describes the fluctuations is continuous due to the macroscopic nature of the variable. On the other hand, it is not evident whether or not the Markov property is a reasonable assumption. This is due to the fact that, as the frequency of the driving increases, the contributions due to the higher order time-derivative terms can become increasingly significant. Hence these higher order terms may have to be accommodated in effecting the stochastic process. Essentially, we could include certain additional variables, apart from the ones that are relevant in the absence of driving, and then assume that the extended set of variables follows continuous Markov process and that its asymptotic distribution describes the oscillating state.
	
	The study of an underdamped Brownian motion, subjected to periodic driving, can be of considerable interest for multiple reasons. First, it is a prototypical example of the Langevin dynamics. Since any continuous Markov process is represented by a Langevin equation, it may be possible to draw useful analogies between the macroscopic variables that follow such a process and the position and velocity variables of the Brownian particle. Second, the velocity variable can be thought of as an additional degree of freedom that is relevant when driving is introduced to the overdamped Brownian motion. The extent of the relevance, for instance, can be estimated by comparing the marginal distribution of the underdamped motion, wherein the velocity is eliminated, with the distribution of the overdamped motion. In the special case of linearly driven Langevin equation, the correlation between the position and velocity degrees can become significant, depending on the driving~\cite{Awasthi2020}. Third, we can track almost analytically the underdamped Brownian motion even in the presence of driving, and hence the study may prove useful in finding some generic features of the oscillating state.
	
	In an earlier work~\cite{Awasthi2020}, we found that any linearly driven Langevin equation can be solved exactly upon exploiting the underlying $SL_2$~symmetry. The exact solution could reveal the presence of oscillating states under certain conditions, could expose interesting properties of some observables and even could establish certain relations among them. This motivates us to extend the enquiry beyond linear driving and search for solutions, if possible exactly, in the presence of anharmonic perturbations. It is also natural to wonder whether we could capitalize on the symmetry even in the nonlinear case. We essentially ask the following central questions. Do driven Langevin systems with anharmonic perturbations reach an oscillating state asymptotically? Can we find the probability distributions of the oscillating state exactly? What are the conditions under which these oscillating states exist and are stable?

	The layout of the current work is as follows. In the next section, we consider a generic class of driven nonlinear Langevin equations and formulate a perturbation scheme wherein the underlying $SL_2$~symmetry is not only manifest but also can be exploited to obtain the corresponding asymptotic distributions to any order. In Sec.~\ref{stability}, we explore the conditions under which the driven systems reach an oscillating state and can remain stable under perturbations. Finally, we summarize and conclude briefly with comments  and remarks in Sec.~\ref{conc}.

	\section{Driven underdamped Brownian particle: Asymptotic distribution}\label{Bparticle}
	
	We will assume that the dynamics of a Brownian particle, when subjected to periodic driving, is governed by the Langevin equation of an underdamped Brownian motion with time-dependent parameters that are $T$-periodic. If we view the Langevin dynamics as an effective description, that is obtained upon eliminating the bath degrees of freedom, then it is reasonable to assume that both viscous coefficient and the noise strength will have to vary with the same period. Though there is no compelling reason for the external potential to vary commensurably, we nevertheless will restrict all time-dependent modulations to the same period. In this section, we shall obtain the asymptotic distribution of the periodically driven underdamped Brownian particle, by perturbatively treating the nonlinear part of the external potential, while nonperturbatively accounting the periodic time dependence.
	
	\subsection{Driven nonlinear Langevin dynamics}\label{Langevin-dynamics}
	
	We shall consider the stochastic dynamics of the position~$X_t$ and velocity~$V_t$ of a driven Brownian particle, governed by the following set of equations,
	\begin{align}\label{stoc-dyn}
		\dot{X}_t& = V_t ~,\nonumber \\
		\dot{V}_t& = -\gamma V_t + f \left( X_t, \lambda \right) + \eta(t)~,
	\end{align} 
	where the viscous coefficient~$\gamma$, the set of parameters~$\lambda$ of the external force~$f$ and the Gaussian noise~$\eta$ have $T$-periodic time dependence. The noise is assumed to have zero mean and nonzero variance~${\langle \eta\tm \eta(t') \rangle_{\eta}= 2D(t) \delta(t-t')}$, where the time-dependent diffusion coefficient~$D$ is $T$-periodic. The external force~$f$ may also contain nonlinear terms of~$X_t$.
	
	The associated Fokker-Planck (FP) equation for the probability distribution~$P(x,v,t)$ of the above nonlinear Langevin dynamics is given by
	\begin{equation}\label{FP-eqn}
		\frac{\partial}{\partial t}P(x,v,t) =\mathcal{L}(x,v, g(t)) P(x,v,t)~,
	\end{equation}
	where~$g$ denotes all the parameters, including~$\gamma$ and~$D$, and the FP operator~$\mathcal{L}$ is defined as
	\begin{equation}\label{FP-op}
		\mathcal{L}(x,v, g) :=  -\frac{\partial}{\partial x}v - \frac{\partial}{\partial v}\left[ -\gm  v +  f(x,\lambda) \right]  + D  \frac{\partial^{2}}{\partial v^{2}} ~.
	\end{equation}
	It should be remarked that though it is common to assume the form of the external force to be~${f=f(x,\lambda)}$, the analysis that we shall employ to obtain the asymptotic distribution is oblivious to this restriction and in fact holds good for~${f=f(x,v,\lambda)}$. Essentially, it is sufficient to assume that~$f$ is Taylor expandable in $x,v$~variables and thus can be written as~${f = - \sum_{n,m} \lambda_{n,m} x^n v^m}$, where~${n,m}$ are non-negative integers, and that the time dependence of the external force is implicitly provided by the linear parameter~$k$ and the nonlinear parameters~${\lambda_{n,m}}$. We will tune off the parameter~$\lambda_{0,1}$ to zero, since the linear term~$\gamma v$ is already included, and denote the linear coefficient~$\lambda_{1,0}$ as~$k$, for notational familiarity.
	
	The asymptotic distribution of the FP equation for periodically driven Brownian particle can be expected to be $T$-periodic under certain conditions. In other words, the driven stochastic system can exist in a state, that we refer to as an oscillating state, wherein various relevant observables exhibit periodic properties. Suppose we consider the domain~$\mathcal{D}_{eq}$ in the parameter space~$\gamma, D, k$ and~$\lambda_{n,m}$ which is defined as the set of all points for which the asymptotic distribution of the FP equation, when the parameters are time independent, is an equilibrium distribution. Let us consider the domain~$\mathcal{D}_{eq}$ to be simply connected and the equilibrium state therein to be continuous. If we now drive the parameters of the FP equation in this domain continuously, with a period much larger than the corresponding equilibrium relaxation timescales, then the asymptotic state essentially will be a periodic trajectory in the equilibrium state space. As we decrease the period, the trajectories of course will not be restricted to the equilibrium state space but are likely to be continuous in some extended space, wherein the closed trajectories presumably can characterize the oscillating states of a driven system. While it is far from obvious what this extended space is or how to characterize the oscillating states, we shall see that it may still be possible to specify some of the conditions under which these states can exist.
	
	The asymptotic distribution is of course obtained by taking the large time limit of the solution of the FP equation~\cite{Awasthi2020}. We could choose the prescription for approaching the limit in periodically driven systems, for instance, by first decomposing time,~${t = NT + \tau}$, in terms of an integer~$N$ number of periods and a remainder~$\tau$ such that~$0 \le \tau < T$, and then by taking~$N \to \infty$ limit. Thus the large time limit of the solution 
	\begin{equation}\label{form-sol}
		P(x, v, t) = \mathcal{U}(x, v; t) P(x, v, 0) ~,
	\end{equation}
	of the FP equation~\eqref{FP-eqn}, results in a $T$-periodic asymptotic distribution 
	\begin{equation}\label{asymp-sol}
		P_{os}(x, v, \tau) =  \mathcal{U}(x, v; \tau) P_{\infty}(x, v, 0)~,
	\end{equation}
	provided the time-independent asymptotic distribution~$P_{\infty}(x, v, 0)$ exists. In the above equation,~$\mathcal{U}$  denotes the evolution operator, which is formally expressed as
	\begin{equation}\label{ker}
		\mathcal{U}(x, v; t) := \mathcal{T}\big\lbrace e^{\int_{0}^{t} \mathcal{L}(x, v, g(t)) dt} \big\rbrace  ~,
	\end{equation}
	where~$\mathcal{T}$ indicates time ordering, and the time-independent distribution
	\begin{equation}\label{largeN-asym}
		P_{\infty}(x, v, 0) := \lim_{N\to\infty} \left[ \mathcal{U}(x, v; T) \right]^N  P(x, v, 0)~,
	\end{equation}
	where~$P(x, v, 0)$ is the initial distribution. The existence and uniqueness of~${P_{\infty}(x, v, 0)}$ is dictated by the spectrum of the monodromy operator~${\mathcal{U}(x, v; T)}$. In case there exists a unique distribution then it necessarily satisfies the eigenvalue equation 
	\begin{equation}\label{t-indep-asym}
		\mathcal{U}(x, v; T)P_{\infty}(x, v, 0)=P_{\infty}(x, v, 0)~.
	\end{equation}
	When the parameters are chosen from the domain~$\mathcal{D}_{eq}$ with their values kept fixed in time, then the above equation is equivalent to~${\mathcal{L} P_{\infty} =0}$ and the asymptotic distribution is an equilibrium distribution~${P_{eq}(x, v)}$. Hence we could equivalently define the domain~$\mathcal{D}_{eq}$ as the set of all~$g$ for which the FP operator~$\mathcal{L}(x,v, g)$ is negative semi-definite, with a unique normalizable eigenfunction corresponding to the zero eigenvalue, and with a nonzero gap between zero and the real part of any other eigenvalue. These properties of the FP operator are also required for the equilibrium state to avoid any singular behavior in~$\mathcal{D}_{eq}$.  Suppose we now drive the parameters~$g=g(t)$ piecewise continuously, though not necessarily respecting the negative semi-definite property of~$\mathcal{L}(x,v, g(t))$, but necessarily ensuring that the real part of none of the eigenvalues of~$\mathcal{U}(x, v; T)$ exceeds unity. Then under such conditions the time-independent asymptotic distribution~${P_{\infty}(x, v, 0)}$ can exist, and hence the system can reach an oscillating state described by the distribution~$P_{os}(x, v, \tau)$. In some sense, the domain~$\mathcal{D}_{os}$ of the parameter space for which the oscillating states exist can be larger than the domain~$\mathcal{D}_{eq}$ for which the equilibrium states exist. This may even suggest that the stability of the states of macroscopic systems can presumably be controlled and manipulated by periodic driving.
	
	In order to comprehend the role of driving in the oscillating states, it is natural to ask the following questions. What are the solutions of the FP equation for a generic driving~$g(t)$? Do these solutions have a large time limit? Namely, does the system reach an oscillating state independent of the initial distribution? Can we specify the necessary conditions that guarantee an oscillating state without having to solve explicitly the FP equation? In other words, can we find some criteria that tells us whether or not a given~$g(t)$ is in the domain~$\mathcal{D}_{os}$? In case of linear external force, using certain techniques from representation theory, we could find almost exactly the oscillating state and establish rigorously the necessary conditions for its existence~\cite{Awasthi2020}.  We shall now show, even in cases where the external forces contain nonlinear terms, that similar techniques can be employed to answer these questions, provided we deal with nonlinearity perturbatively.

	\subsection{Perturbative analysis to all orders}\label{pert-theory}
	
	Following the standard perturbative analysis, we can decompose the FP operator,
	\bee
	\mathcal{L}=\mathcal{L}_0+ \epsilon \mathcal{L}_p~,
	\eee
	into a solvable part that includes only the linear force terms, given by
	\bee
	\mathcal{L}_0 = \mathcal{L}_0 (x, v, g_0) = -\frac{\partial}{\partial x}v + \frac{\partial}{\partial v} \left[ \gm  v +  kx \right]  + D  \frac{\partial^{2}}{\partial v^{2}}~,
	\eee
	where~$g_0$ denotes~$\gamma, D$ and~$k$, and a perturbative part
	\bee
	\mathcal{L}_p = \mathcal{L}_p (x, v, \lambda)  :=  \frac{\partial}{\partial v} \mathcal{O}_p~,
	\eee
	where~$\mathcal{O}_p = \mathcal{O}_p(x, v, \lambda) = -kx -  f(x, v, \lambda)$ consists only the nonlinear terms of the force~$f$.
	The dimensionless parameter~$\epsilon$ is introduced to conveniently track the order of nonlinearity. Expanding the probability distribution~$P(x, v, t)$, in the FP equation~\eqref{FP-eqn}, as the following series in the perturbative parameter~$\epsilon$,
	\bee
	P(x, v, t) := \sum_{n=0}^{\infty} \epsilon^n P^{(n)}(x, v, t)~,
	\eee
	leads to a sequence of dynamical equations for~$P^{(n)}$. 
	
	The zeroth-order equation is given by
	\bee\label{zero-FP-eq}
	\frac{\partial}{\partial t}P^{(0)}(x,v,t) =\mathcal{L}_0(x,v, g_0(t)) P^{(0)}(x,v,t)~,
	\eee
	which is the FP equation of a driven Brownian particle in harmonic potential~\cite{Awasthi2020}. The other higher-order corrections~$P^{(n)}$ are governed by the recursive equations, 
	\bee\label{nth-FP-eq}
	\frac{\partial}{\partial t}P^{(n)} =\mathcal{L}_0 P^{(n)} + \mathcal{L}_p P^{(n\!-\!1)}~,
	\eee
	for any integer~$n \ge 1$. The dependence on the coordinates~$x, v, t$ and on other parameters is not explicitly exhibited here for notational simplicity.
	
	We will assume~$P(x, v, t)$ is normalized to any given order of~$\epsilon$. In other words,~$P^{(0)}(x, v, t)$ is normalized, and~$\int dx dv P^{(n)}(x, v, t) =0$ for all~$n \ge 1$. We denote the asymptotic perturbative components of~$P(x, v, t)$ as~$P^{(n)}_{\infty}(x, v, t)$, which are obtained by taking the large time limit of the solutions of Eqs.~\eqref{zero-FP-eq} and~\eqref{nth-FP-eq}. We could choose the initial conditions for these equations such that the initial zeroth-order distribution~${P^{(0)}(x, v, 0)= P(x,v,0)}$, and hence all the initial higher-order corrections~${P^{(n)}_{\infty}(x, v, 0)=0}$, for~${n \ge 1}$.
	
	Let us consider the driving parameters such that the necessary conditions required for the solution of Eq.~\eqref{zero-FP-eq} to asymptotically approach a well-defined limit are satisfied. The asymptotic distribution~$P^{(0)}_{\infty}(x,v,t)$ is in fact a Gaussian distribution with zero mean and  a $T$-periodic covariance matrix
	\bee\label{cov-mat}
	\Sigma(t) := 
	\begin{bmatrix}
		\langle x^2 \rangle_0 & \langle xv \rangle_0 \\
		\langle vx \rangle_0 &\langle v^2 \rangle_0
	\end{bmatrix}
	\equiv
	\begin{bmatrix}
		\widetilde{X}_{2,0}(t) & \widetilde{X}_{1,1}(t)\\
		\widetilde{X}_{1,1}(t) & \widetilde{X}_{0,2}(t)
	\end{bmatrix}~,
	\eee
	whose matrix elements depend on the parameters~$g_0(t)$.
	The methods employed to find the conditions for the existence of the asymptotic limit and to obtain the distribution~$P^{(0)}_{\infty}(x,v,t)$ are detailed in the Ref.~\cite{Awasthi2020}. Since we shall extend these methods to obtain the higher-order asymptotic functions~${P^{(n)}_{\infty}(x, v, t)}$, for~${n \ge 1}$, we will briefly spell them out, as and when required. 
	
	The algorithm that we shall employ to evaluate the asymptotic $n$-th order correction is as follows. We first substitute~$P^{(n-1)}$  in the equation~\eqref{nth-FP-eq} with its asymptotic function~${P^{(n-1)}_{\infty}}$ which we assume exists,  and then find the solution of the modified equation. This solution is valid only for times that are large compared to the time required for the $(n-1)$th-order distribution to reach its asymptotic limit. In other words, decompose the time~${t= (N_0 + N_1)T + \tau}$ and solve Eq.~\eqref{nth-FP-eq} for~${t= N_1T + \tau}$ with the new initial condition given at time~$N_0T$, where~$N_0$ is chosen such that the substitution~$P^{(n-1)}$ with~${P^{(n-1)}_{\infty}}$ is justifiable, and then take the limit~${N_1 \to \infty}$. We can then show that the asymptotic limit~${P^{(n)}_{\infty}(x, v, t)}$ of the solution to the modified Eq.~\eqref{nth-FP-eq} exists when a certain condition holds for any arbitrary initial function~${P^{(n)}(x, v, N_0T)}$. We shall discuss later and analyse in detail this specific condition which is in fact independent of~$n$. Thus when this condition holds and the zeroth-order distribution~$P^{(0)}_{\infty}(x,v,t)$ exists, we establish iteratively the existence of~${P^{(n)}_{\infty}(x, v, t)}$ for any~$n >1$. 
	
	The calculations for determining explicitly the  $n$-th order correction is similar to those for determining the first-order correction.  We begin by rewriting the first-order correction
	\bee\label{P1-A1}
	P^{(1)}(x,v,t) = - \left( A^{(1)} - \langle A^{(1)} \rangle_0 \right)P^{(0)}_{\infty}(x,v,t)~,
	\eee
	where~$A^{(1)}=  A^{(1)}(x,v,t)$ is yet to be determined function and~$\langle A^{(1)} \rangle_0$ is the average of~$A^{(1)}$ with respect to~$P^{(0)}_{\infty}(x,v,t)$. If the perturbation~$\mathcal{O}_p(x, v, \lambda)$ is a polynomial function in~$x$ and~$v$ then~$A^{(1)}(x,v,t)$ will also be a polynomial in these variables. Owing to the underlying $SL_2$~symmetry of the unperturbed FP equation, it is beneficial  to choose the basis functions
	\bee
	\mathcal{O}^r_L := x^{L-r} v^r~,
	\eee
	where~$L$ is a positive integer and denotes the degree of homogeneity and~$r$, for a given~$L$, runs over the integers from~$0$ to~$L$. Let us consider the perturbation~$\mathcal{O}_p$ to be a nonlinear polynomial of degree~$L_p$, and hence can be written in the form
	\bee\label{Op-poly}
	\mathcal{O}_p(x, v, \lambda) = \sum_{L=2}^{L_p} \sum_{r=0}^L \lambda^L_r \mathcal{O}^r_L~,
	\eee
	where the parameters~$\lambda^L_r$, in case they depend on time, are~$T$-periodic. We can represent the function~$A^{(1)}$ in this basis as
	\bee\label{A1-poly}
	A^{(1)}(x, v, t) =  \sum_{L=1}^{L_1} \sum_{r=0}^L   a^L_r(t) \mathcal{O}^r_L~,
	\eee
	where the coefficients~$a^L_r(t)$ are time dependent and~$L_1$ is a finite positive integer such that~${L_1 \gg L_p}$.  We shall henceforth refer the coefficients of~$\mathcal{O}^r_L$ in an expansion as level-$L$ coefficients. 
	
	We now proceed to determine the coefficients~$a^L_r(t)$. Substituting Eqs.~\eqref{P1-A1},~\eqref{Op-poly} and~\eqref{A1-poly} in the recursive equation~\eqref{nth-FP-eq}, for~$n=1$, straightforwardly leads to a polynomial equation. Then equating the coefficients of each monomial in the polynomial equation results in an ordinary differential equation for the variables~$a^L_r$. These dynamical equations can be written in the form
	\begin{equation}\label{a-dyn}
		\frac{d}{dt} a^L_r = H^L_r + N^L_r  + R^L_r + S^L_r  ~,
	\end{equation}
	where the first term~$H^L_r$ contains only the level-$L$ coefficients of~$A^{(1)}$ and is given by
	\begin{equation}\label{hom-def}
		H^L_r = - (L+1-r) a^L_{r-1} + r \gamma_p a^L_r  + (r+1) k_p  a^L_{r+1} ~,
	\end{equation}
	and include the $T$-periodic parameters
	\begin{eqnarray}\label{gp-kp}
		\gamma_p := \gamma - 2 D (\Sigma^{-1})_{22}~, \nonumber \\
		k_p := k - 2 D (\Sigma^{-1})_{12}~;
	\end{eqnarray}
	the second term~$N^L_r$ contains only a level-$(L+2)$ coefficient of~$A^{(1)}$ and is given by
	\begin{eqnarray}\label{nhom-def}
		N^L_r =  \begin{cases}
			D (r+1)(r+2) a^{L+2}_{r+2}~, \text{ for  } {1 \le L \le L_1 -2}~, \\ 
			0~, \text{ for  }  L \ge L_1-1~;
		\end{cases}
	\end{eqnarray}
	the third term~$R^L_r$ contains only a level-$(L+1)$ coefficient of the perturbation~$\mathcal{O}_p$ and is given by
	\begin{eqnarray}\label{pert1-def}
		R^L_r =  \begin{cases}
			- (r+1) \lambda^{L+1}_{r+1}~, \text{ for  } {1 \le L \le L_p -1}~, \\ 
			0~, \text{ for  }  L \ge L_p~;
		\end{cases}
	\end{eqnarray}
	while the fourth term contains only level-$(L-1)$ coefficients of the perturbation~$\mathcal{O}_p$ and is given by
	\begin{eqnarray}\label{pert2-def}
		S^L_r =  \begin{cases}
			\Sigma^{-1}_{12} \lambda^{L-1}_{r} + \Sigma^{-1}_{22} \lambda^{L-1}_{r-1}~, \text{ for  } {3 \le L \le L_p +1}~, \\ 
			0~, \text{ for  either } { L \ge L_p +2}~\text{ or } {1 \le L \le 2}~.
		\end{cases}
	\end{eqnarray}
	Furthermore, the constant term of the polynomial equation leads to an additional equation
	\begin{equation}
		\frac{d}{dt} \langle A^{(1)} \rangle_0  + 2 D a^2_2 =0~,
	\end{equation}
	which though is not an independent one. Note that the dynamical equations~\eqref{a-dyn} are such that the level-$L$ coefficients~$a^L_r$ are all linearly coupled amongst themselves. Further they contain the inhomogeneous terms involving both the variables that are level-${(L+2)}$ coefficients~$a^{L+2}_{r+2}$ and the given interaction terms~$ R^L_r$ and $S^L_r $. The interaction terms though are absent  for the dynamical equations with~${L \ge L_p + 2}$. Hence it is evident that we need to solve for~$a^{L+2}_{r+2}$ in order to solve for~$a^L_r$.
	
	In order to solve Eqs.~\eqref{a-dyn} which are inhomogeneous, it is useful to first discuss the symmetries and solutions of the corresponding homogeneous equations obtained from Eqs.~\eqref{a-dyn} by dropping $N^L_r $, $ R^L_r $ and $ S^L_r $ terms. The homogeneous equations can be rewritten as
	\begin{equation}\label{a-hom}
		\frac{d}{dt} {\bf a}_L=  \left[ - \mathbf{J}^{-}_{L} +\frac{\gamma_p }{2} \left( L  I_L +\mathbf{J}_{L} \right) + k_p  \mathbf{J}^{+}_{L}\right]^T {\bf a}_L~,
	\end{equation}
	where~${\bf a}_L$ is a $(L+1)$-component vector whose elements are~$a^L_r$, namely~${\bf a}_L^T := \left( a^L_0, a^L_1, \cdots, a^L_L \right)$, the superscript~$T$  on a vector or a matrix denotes their transpose, the matrix~$I_L$ is the identity matrix of dimension~$(L+1)$, and~$\lbrace \mathbf{J}^{\pm}_{L},\mathbf{J}_{L} \rbrace$ are matrices of the same dimension with matrix elements 
	\begin{eqnarray}
		\left( \mathbf{J}^{+}_{L}  \right)_{r,s}&=& r \delta_{r,s\!+\!1} ~,\nonumber \\
		\left( \mathbf{J}^{-}_{L}  \right)_{r,s}&=& (L-r) \delta_{r,s\!-\!1}~, \nonumber  \\
		\left( \mathbf{J}_{L}  \right)_{r,s}&=& (2r-L) \delta_{r,s}~,
	\end{eqnarray}
	and whose indices~${r,s}$ run over all the integers from~$0$ to~$L$. We can easily verify that these matrices satisfy the commutation relations of the $sl_2$~algebra, namely
	\bee\label{sl2alg}
	\left[ \mathbf{J}_{L} , \mathbf{J}^{\pm}_{L} \right] = \pm 2\mathbf{J}^{\pm}_{L} ~,\quad  \left[\mathbf{J}^{+}_{L},  \mathbf{J}^{-}_{L} \right] =  \mathbf{J}_{L} ~.
	\eee
	In fact these matrices form an irreducible representation of the generators of the group~$SL_2(R)$.
	This dynamical symmetry exhibited by the homogeneous equations is indeed induced by the underlying $SL_2$~symmetry of the unperturbed FP operator.
	
	It may be remarked in passing that Eq.~\eqref{a-hom} can be mapped to a transposed version by a similarity transformation induced by an anti-diagonal matrix~${\mathcal S}$ with matrix elements~${\mathcal S}_{r,s} = \delta_{r+s,L} r! (L-r)!/L!$. Under this transpose map $\mathbf{J}^{+}_{L}$ and $\mathbf{J}^{-}_{L}$ reverse their role in the algebra  as is evident from the relations ${\mathcal S}^{-1} \mathbf{J}_{L}{\mathcal S} = -\left( \mathbf{J}_{L} \right)^T$ and
	${\mathcal S}^{-1} \mathbf{J}^{\pm}_{L}{\mathcal S} = \left( \mathbf{J}^{\pm}_{L} \right)^T$.
	
	The explicit $L$-dependent term in Eq.~\eqref{a-hom} can be removed by writing the vector~${\bf a}_L$ in terms of another vector~${\bf b}_L = {\bf a}_L \exp(-L\Gamma_p/2)$, where~$\Gamma_p(t)= \int_o^t dt' \gamma_p(t')$, and thus obtain the equation 
	\begin{equation}\label{b-hom}
		\frac{d}{dt} {\bf b}_L=  \left[ - \mathbf{J}^{-}_{L} +\frac{\gamma_p }{2} \mathbf{J}_{L} + k_p  \mathbf{J}^{+}_{L}\right]^T {\bf b}_L~,
	\end{equation}
	which has a form independent of any specific $sl_2$~representation. This form makes it amenable to determine its solutions by exploiting  the theorem which states that any irreducible representation of~$sl_2$ is a symmetric power of the standard representation~\cite{Fulton2004}. In other words, we obtain the solutions of Eq.~\eqref{b-hom} by taking the symmetrized tensor products of the solutions of its corresponding equation in the standard representation, which is 
	\begin{equation}\label{b-hom1}
		\frac{d}{dt} {\bf b}_1=  \left[ - \mathbf{J}^{-}_{1} +\frac{\gamma_p }{2} \mathbf{J}_{1} + k_p  \mathbf{J}^{+}_{1}\right]^T {\bf b}_1~,
	\end{equation}
	or, when written in terms of components of the vector~${\bf b}_1^T := (b_0, b_1)$, is
	\begin{align}\label{standard-b}
		\frac{d}{dt}
		\begin{pmatrix}  
			b_0\\ 
			b_1 \\ 
		\end{pmatrix} = 
		\begin{pmatrix}
			- \frac{1}{2}\gamma_p & k_p \\
			-1 & \frac{1}{2}\gamma_p 
		\end{pmatrix}
		\begin{pmatrix}  
			b_0 \\ 
			b_1 \\ 
		\end{pmatrix} ~.
	\end{align}  
	Notice that the component~$b_1$ satisfies the Hill equation
	\begin{equation}\label{hill-eq}
		\frac{d^2}{dt^2} b_1 + \nu_p b_1 =0~,
	\end{equation}
	where the $T$-periodic parameter
	\begin{equation}\label{nu-p}
		\nu_p = k_p -  \frac{1}{2} \dot{\gamma}_p - \frac{1}{4} \gamma_p^2~.
	\end{equation}
	Hence the solutions of Eqs.~\eqref{b-hom1}, \eqref{b-hom} and \eqref{a-hom} can be solely expressed in terms of the two independent solutions of the Hill equation, denoted~$u(t)$ and~$w(t)$, which can be chosen to be pseudo-periodic with Floquet exponents~$\mu_p$ and~$-\mu_p$, respectively, namely~${u(t+T)=u(t) \exp(\mu_p T)}$ and ~${w(t+T)=w(t) \exp(-\mu_p T)}$. The psuedo-periodic solutions of Eq.~\eqref{b-hom1} are
	\begin{align}\label{sol-b1}
		\mathbf{b}_{1}^{(0)} = 
		\begin{pmatrix}  
			\frac{1}{2} \gamma_p u-\dot{u}  \\ 
			u\\ 
		\end{pmatrix} ~,~
		\mathbf{b}_{1}^{(1)} = 
		\begin{pmatrix}  
			\frac{1}{2} \gamma_p w - \dot{w} \\ 
			w\\ 
		\end{pmatrix}~,
	\end{align}
	while those of the homogeneous equation~\eqref{a-hom} are
	\begin{equation}\label{sol-bL}
		\mathbf{a}_{L}^{(r)} = e^{\frac{L}{2} \Gamma_p} \text{Sym} \left[ \left( \mathbf{b}_{1}^{(0)} \right)^{\otimes(L-r)} \otimes \left(  \mathbf{b}_{1}^{(1)} \right)^{\otimes r} \right]~,
	\end{equation}
	where~$r$ runs over the integers from~$0$ to~$L$ and $\text{Sym}$ denotes the symmetrization of the tensor products of the vectors in the bracket. The solutions $ \mathbf{a}_{L}^{(r)}$ are pseudo-periodic with corresponding Floquet exponents
	\begin{equation}\label{Fexp-L}
		\mu_L^{(r)} = \frac{1}{2}L\overline{\gamma}_p + (L-2r)\mu_p~,
	\end{equation}
	where~$\overline{\gamma}_p$ is the time average of~$\gamma_p$ over a period~$T$, and they vanish in the large-time limit provided the modulus of the real part of the fundamental Floquet exponent
	\begin{equation}\label{cond-pert}
		\left| \text{Re} \left( \mu_p \right) \right| < -\frac{1}{2} \overline{\gamma}_p~.
	\end{equation}

	Henceforth we will assume that the parameters~${k, \gamma}$ and~$D$  are chosen such that the condition~\eqref{cond-pert} holds. We can now deduce the large-time solutions~$a^L_r(t)$ of Eq.~\eqref{a-dyn} for each~$L$ sequentially in the reverse order starting from~$L=L_1$ down to~$L=1$. For~${L_p+2 \le L \le L_1}$, the set of equations~\eqref{a-dyn} are homogenous and hence the corresponding large-time solutions~$a^L_r(t)$ vanish for any arbitrary initial conditions. The equations~\eqref{a-dyn} for~$L=L_p+1$ and~$L=L_p$ are though inhomogeneous  have only given $T$-periodic inhomogeneous terms~$ h^L_r =R^L_r + S^L_r$. The solutions of these equations can formally be written as
	\begin{equation}\label{a-L-soln}
		\mathbf{a}_L(t)
		= K_L(t,0) \mathbf{a}_L(0)
		+ \int_{0}^{t} ds K_L(t,s)\mathbf{h}_L(s)~,
	\end{equation}
	where for any~$L$ the vector~$\mathbf{h}_L$ is defined by the components~$ h^L_r$, and the matrix
	\begin{equation}
		K_L(t,s)= \Phi_L(t)\Phi^{-1}_L(s)
	\end{equation}
	is defined by the fundamental matrix~$\Phi_L(t)$ of Eq.~\eqref{a-hom} constructed with~$ \mathbf{a}_{L}^{(r)}(t)$ as column vectors. Now to determine the large-time limit of the solutions, we use the following two properties of the matrix~$K(t,s)$. First, the matrix can be decomposed as~${K_L(t,s)= K_L(t,t')K_L(t',s)}$ for any~$t'$, which is a consequence of its definition. Second, it is invariant under discrete time translation by a period, namely ${K_L(t+T,s+T)= K_L(t,s)}$, which is due to the pseudo-periodic nature of the fundamental matrix,~${\Phi_L(t+T) = \Phi_L(t) \Lambda_L}$, where~$\Lambda_L$ is a diagonal matrix with elements~$\exp(\mu_L^{(r)}T)$. Using these properties, for any time~$t= NT + \tau$ the first term on the right hand side of Eq.~\eqref{a-L-soln} can be written as 
	\begin{equation}\label{term1}
		K_L(\tau,0) \left[ K_L(T,0)\right]^N \mathbf{a}_L(0)~,
	\end{equation}
	while for any $T$-periodic function~$\mathbf{h}_L(t)$ the second term can be written as
	\begin{equation}\label{term2}
		K_L(\tau,0) \left( 1 \!-\! K_L(T,0)^N \right)  Z_L(T;\mathbf{h}_L) + Y_L(\tau;\mathbf{h}_L)~,
	\end{equation}
	where the matrices~$Y_L(\tau;\mathbf{h}_L)$ and~$Z_L(T;\mathbf{h}_L)$ are independent of~$N$ and are defined for any given $T$-periodic vector~$\mathbf{f}_L(t)$ as
	\begin{eqnarray}\label{termYZ}
		Y_L(\tau;\mathbf{f}_L ) &=&  \int_{0}^{\tau} \!ds K_L(\tau,s)\mathbf{f}_L(s)~,\nonumber \\
		Z_L(T; \mathbf{f}_L) &=& \left({ 1- K_L(T,0) }\right)^{-1}Y_L(T;\mathbf{f}_L )~.
	\end{eqnarray}
	Note that~$K_L(T,0) = \Phi_L(0) \Lambda_L \Phi^{-1}_L(0)$, and hence its eigenvalues are same as those of~$\Lambda_L$, namely~$\exp(\mu_L^{(r)}T)$. Now when the condition~\eqref{cond-pert} holds not only the matrix~$Z_L(T;\mathbf{h}_L)$ is ensured to be nonsingular but also the matrix~$K_L(T,0)^N$ approaches zero in the limit~${N \to \infty}$. Consequently, a well-defined asymptotic limit of the solution that is independent of the initial conditions exists, and we obtain the asymptotic solution as
	\begin{equation}\label{asy-sol-L}
		\mathbf{a}_L(\tau) = K_L(\tau,0) Z_L(T;\mathbf{h}_L )+Y_L(\tau;\mathbf{h}_L )~,
	\end{equation}
	for~$L=L_p+1$  and~$L=L_p$. It is straightforward to verify that ${\mathbf{a}_L(\tau+T)= \mathbf{a}_L(\tau)}$. This also implies that~$N^L_r(t)$ becomes $T$-periodic for~$L=L_p-1$ and~$L=L_p-2$. It should emphasized that the asymptotic solution~\eqref{asy-sol-L} is $T$-periodic in spite of the psuedo-periodic nature of the homogeneous part of the solution~\eqref{a-L-soln}. 
	
	By hierarchically repeating the arguments that lead to Eq.~\eqref{asy-sol-L}, we further obtain the other $T$-periodic asymptotic solutions 
	\begin{equation}\label{asy-sol-L-2}
		\mathbf{a}_L(\tau) = K_L(\tau,0) Z_L(T;\mathbf{h}_L+ \mathbf{n}_L )+Y_L(\tau;\mathbf{h}_L+\mathbf{n}_L )~,
	\end{equation}
	for $1 \le L \le L_p -1$, where the components of the vector~$\mathbf{n}_L$ are defined to be the asymptotic values of~$ N^L_r$.
	
	We can of course determine the next-order correction~${P^{(2)}_{\infty}(x, v, t)}$ by exactly going through the same mathematical manipulation as performed earlier to determine~${P^{(1)}_{\infty}(x, v, t)}$ after replacing the nonlinear term~$\mathcal{O}_p$ with~$ ( \langle A^{(1)} \rangle_0 - A^{(1)} )\mathcal{O}_p$. Thus proceeding iteratively,  correction term~${P^{(n)}_{\infty}(x, v, t)}$ to any order~$n$ can be obtained.
	
	To summarize this section, we have shown that the asymptotic probability distribution of the periodically driven nonlinear FP equation~\eqref{FP-eqn} to any order of perturbation in anharmonicity exists and is $T$-periodic provided the unperturbed oscillating state exists and the condition~\eqref{cond-pert} holds.  Furthermore all the coefficients of this $T$-periodic asymptotic distribution to any order can be determined exactly in terms of the solutions of the Hill equation~\eqref{hill-eq}.

	\section{Stability of the Oscillating state}\label{stability}
	
	In this section, we study the stability of the oscillating states and the effect of the perturbations. In particular, we will survey the domains of the unperturbed oscillating states and then examine whether perturbations can coexist within these domains. 
	
	\subsection{Unperturbed oscillating states}
	
	The zeroth-order asymptotic distribution~$P^{(0)}_{\infty}(x,v,t)$ exists when the first moments vanish at large times and the asymptotic covariance matrix is positive definite~\cite{Awasthi2020}. The first moments of the distribution~$P^{(0)}(x,v,t)$ are related to the solutions~$Y_{10}$ of the following Hill equation
	\begin{equation}\label{hill-eq-0}
		\frac{d^2}{dt^2} Y_{10} + \nu Y_{10} =0~,
	\end{equation}
	where~${\nu = k  - \dot{\gamma}/2 - \gamma^2/4}$. When the Floquet exponents~$\pm \mu$ of this Hill equation satisfy the condition
	\begin{equation}\label{cond-mu-g0}
		| Re{(\mu)} | < \frac{1}{2} \overline{\gamma}~,
	\end{equation} 
	then the first moments vanish asymptotically. This condition also means that ${\exp(-\overline{\gamma} t/2) Y_{10}(t)}$ should remain bounded at all times.  Furthermore the condition~\eqref{cond-mu-g0} ensures that the second moments also remain bounded and become $T$-periodic asymptotically independent of the initial conditions of the distribution. We could deduce this fact since the second moments also can be written in terms of the Floquet solutions of Eq.~\eqref{hill-eq-0}. Essentially the dynamical equations of the second moments $X_{2,0}, X_{1,1}$ and $X_{0,2}$ are  
	\begin{eqnarray}\label{mom-2-as}
		\frac{d}{dt}X_{2,0} &=& 2 X_{1,1}~, \nonumber \\
		\frac{d}{dt}X_{1,1} &=&-k X_{2,0}  -\gamma X_{1,1} + X_{0,2}~,\nonumber \\
		\frac{d}{dt}X_{0,2} &=& -2k X_{1,1} -2 \gamma X_{0,2} +2D~,
	\end{eqnarray}
	and the moments $X_{2,0}, X_{1,1}$ and $X_{0,2}$ approach their asymptotic values $\widetilde{X}_{2,0}, \widetilde{X}_{1,1}$ and $\widetilde{X}_{0,2}$, respectively, in the large time limit. The homogeneous part of these equations with parameters $k$ and $\gamma$ has similar structure as Eq.~\eqref{a-hom} for $L=2$ with corresponding parameters $k_p$ and $-\gamma_p$. Hence the manner in which Eq.~\eqref{a-hom}, its solution~\eqref{sol-bL} and its exponents~\eqref{Fexp-L} are related to the Hill equation\eqref{hill-eq} is the exact manner in which the homogeneous part of Eq.~\eqref{mom-2-as}, its solution and its exponents are related to the Hill equation~\eqref{hill-eq-0}. In fact the Floquet exponents associated with the covariance matrix can also be read from Eq.~\eqref{Fexp-L} and are ${-{\overline{\gamma} + (2-2r)\mu}}$, where $r=0,1,2$. 
	
	The condition~\eqref{cond-mu-g0} involves Floquet exponents which depend on the parameters~$k$ and $\gamma$ implicitly, and hence it is far from obvious without explicit verification whether a given driving allows the system to be in an oscillating state. We can rewrite the condition in a form that is more amenable to computation using Floquet theory of the Hill equation. The Floquet exponents essentially can be expressed in terms of the solutions of the Hill equation at time~$T$~\cite{Magnus2013,Eastham1975}. Let $u_1(t)$ and $u_2(t)$ be two independent solutions of the Hill equation~\eqref{hill-eq-0} with the initial conditions:~$u_1(0)=1$, $u_2(0)=0$, $\dot{u}_1(0)=0$ and $\dot{u}_2(0)=1$. Since the parameter~$\nu$ is $T$-periodic,
	\begin{align}\label{sol-T}
		\begin{pmatrix}  
			u_1(t+T)\\ 
			u_2(t+T) \\ 
		\end{pmatrix} = 
		\Phi(T)
		\begin{pmatrix}  
			u_1(t) \\ 
			u_2(t) \\ 
		\end{pmatrix}~,
	\end{align}  
	where the monodromy matrix
	\begin{align}\label{mon-mat}
		\Phi(T) = 
		\begin{pmatrix}
			u_1(T) & \dot{u}_1(T) \\
			u_2(T) & \dot{u}_2(T)
		\end{pmatrix}~.
	\end{align} 
	The Floquet coefficients~$\exp(\pm \mu T)$ are the eigenvalues of the matrix~$\Phi(T)$, namely the roots of the equation ${\rho^2 - \rho \Delta + 1 =0}$, where the trace of the matrix
	\begin{equation}\label{tr-1}
		\Delta = u_1(T) + \dot{u}_2(T)~,
	\end{equation}
	while the determinant is unity since the Wronskian of two independent solutions is constant. In other words, the coefficients
	\begin{equation}\label{floq-coeff}
		e^{\pm \mu T} = \frac{\Delta}{2} \pm \sqrt{ \frac{\Delta^2}{4}-1 }~.
	\end{equation}
	The Floquet coefficients are real for~$|\Delta| \ge 2$ and complex with modulus one for~$|\Delta| < 2$. Substituting Eq.~\eqref{floq-coeff} in condition~\eqref{cond-mu-g0} leads to the relation
	\begin{equation}\label{eff-cond-1}
		|\Delta| < 2\cosh\left( \frac{1}{2} \overline{\gamma} T \right)~.
	\end{equation}
	We can now distinguish different regions. When~${|\Delta| > 2\cosh( \overline{\gamma} T /2)}$ the oscillating states do not exist since the function ${\exp(-\overline{\gamma} t/2) Y_{10}(t)}$ blows up at large time. In the region ${2 < |\Delta| < 2\cosh( \overline{\gamma} T /2)}$, any given initial distribution relaxes into an oscillating state in a time scale $\tau_R \sim  1/ [\overline{\gamma} /2 - | Re{(\mu)} |] $ that depends on the value of~$|\Delta| $.
	While in the region $ |\Delta| \le 2 $, the oscillating states exist and can be reached in time   $\tau_R \sim  2/\overline{\gamma}$. The case ${|\Delta| =2\cosh( \overline{\gamma} T /2)}$ is physically less attractive as it not only requires a fine-tuned driving but also results in an asymptotic state with a memory of the initial conditions.
	
	\begin{figure*}[!htb]
		\centering
		\includegraphics[width=\linewidth]{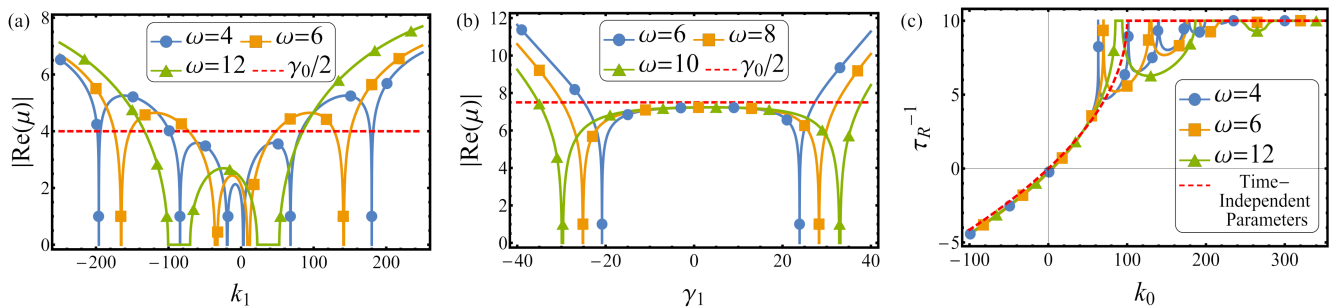}
		\caption{The absolute value of the real part of the Floquet Exponent $|Re(\mu)|$ controls the stability of the system. If its value is below the dotted line of $\gamma_0/2$ (= $\overline{\gamma}/2$) in plot (a) and (b), then the system is asymptotically stable. (a) $|Re(\mu)|$ as function of $k_1$ with $k_0 = 10$, $\gamma_0=8$, and $\gamma_1 = 4$ for frequencies $\omega \in \lbrace 4,6,12 \rbrace$. (b) $|Re(\mu)|$ as function of $\gamma_1$ with $k_0 = 4$, $k_1=2$, and $\gamma_0 = 15$ for frequencies $\omega \in \lbrace 6,8,10 \rbrace$. (c) The relaxation time $\tau_R^{-1}$ as a function of $k_0$  with $k_0 = 4$, $k_1=2$, and $\gamma_0 = 15$ for frequencies $\omega \in \lbrace 6,8,10 \rbrace$. The dotted line denotes the relaxation time when the parameters $k$ and $\gamma$ are taken to be time-independent~($k_1 = \gamma_1 = 0$).  }
		\label{MuPlots}
	\end{figure*}
	
	\begin{figure*}[!htb]
		\centering
		\includegraphics[width=\linewidth]{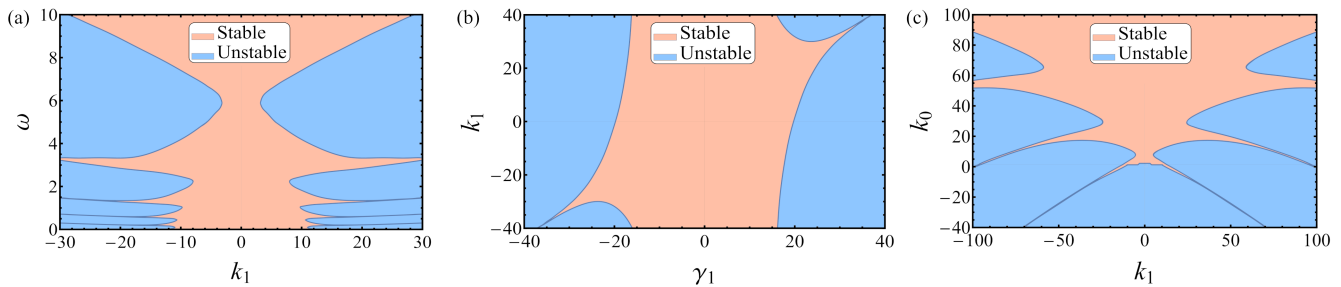}
		\caption{Stable and unstable region for periodically driven system in (a) $k_1-\omega$ plane  with $\gamma_1 = 0$, $k_0 = 10.0625$, and $\gamma_0 = 0.5$;  (b) $\gamma_1-k_1$ plane with $\gamma_0 = 15$, $\omega = 5$, and $k_0 = 15$; (c) $k_1-k_0$ plane with $\gamma_0 = 1$, $\gamma_1 = 0$, and $\omega = 5$.  }
		\label{Stability Region}
	\end{figure*}
	
	In general it may not be possible to analytically determine the explicit dependence of~$\Delta$ on the time-dependent parameters~$k$ and~$\gamma$ or more precisely on the function~$\nu$ of these parameters that appears in the 
	equation~\eqref{hill-eq-0}. Nevertheless we could gain a considerable qualitative understanding of this dependency by probing numerically. To this end we consider the cases where the parameters are restricted to the first harmonics and evaluate~$\Delta$ and in turn determine the relaxation times and depict the domains of the oscillating states. 
	
	Let us drive the parameters as follows,
	\begin{eqnarray}\label{kg-1har}
		&k(t) = k_0 + k_1 \cos(\omega t)~,  \nonumber  \\
		&\gamma(t) = \gamma_0 + \gamma_1 \cos(\omega t)~,
	\end{eqnarray}
	where~$k_0, k_1, \gamma_0, \gamma_1$ and~$\omega=2\pi/T$ are constants. Note that the Hill equation can be viewed as an eigenvalue problem or equivalently as a steady state Schr\"odinger equation of an electron in one-dimensional periodic potentials. The parameter~$k_0$, for instance, can be viewed as an eigenvalue provided the corresponding eigenfunction exist. Unlike the steady state wavefunctions in the electron case, here the eigenfunctions can explode asymptotically though not faster than~$\exp(\gamma_0 t/2)$. As we vary~$k_0$, keeping the other constants fixed, we can expect to encounter both the forbidden regions where the eigenfunctions do not exist and the allowed regions where the eigenfunctions and hence the oscillating states exist. Similar features are indeed expected to show up when the other parameters are varied too.
	
	We find it convenient to plot the modulus of the Floquet exponent~$\mu$ as a function of the parameters in the chosen domain, since $\Delta$ grows exponentially with linear increase of~$\mu$ and hence a log scale can be avoided. The dependence of~$|Re(\mu)|$ for different values of~$\omega$ on~$k_1$ and on~$\gamma_1$ is illustrated  in Fig.~\hyperref[MuPlots]{1(a)} and Fig.~\hyperref[MuPlots]{1(b)}  respectively. All the numerical error bars are essentially less than the thickness of the lines used in the plots. 
	
	As we vary~$k_1$, one passes through three different regions. One of which is where the value of~$|Re(\mu)|$ varies but is less than~$\overline{\gamma}/2$ indicating the system can relax into a stable oscillating state in a time scale that depends on~$k_1$. The other is where~$|Re(\mu)| =0$ and the system can relax to an oscillating state in a time that is independent of~$k_1$. The third kind is where~$|Re(\mu)| > \overline{\gamma}/2$ and the system can not exist in an oscillating state. Note that the region where~$|Re(\mu)|$ approaches zero for $\omega=4$ and~$6$ appear as points in Fig.~\hyperref[MuPlots]{1(a)} only due to the choice of the scale of the axis. The time rate of convergence to an oscillating state can of course be read from Fig.~\hyperref[MuPlots]{1(a)} and Fig.~\hyperref[MuPlots]{1(b)}, but is explicitly plotted in Fig.~\hyperref[MuPlots]{1(c)} to demonstrate the nontrivial dependence of the parameters, say~$k_0$.
	
	It may be required to add the caveat that even though for small~$k_1$ values the relaxation time decreases with increase of driving frequency, this is not a generic feature for arbitrary values of~$k_1$ as is evident from Fig.~\hyperref[MuPlots]{1(a)}. A similar conclusion can be arrived at from Fig.~\hyperref[MuPlots]{1(b)} where the relaxation time is not always monotonically related to the value of~$\gamma_1$. Nevertheless observe in Fig.~\hyperref[MuPlots]{1(b)} that the value of~$\gamma_1$ that saturates the stability of the oscillating state increases with the increase of driving frequency, which is a result that is in tune with the intuition that periodic driving enhances stability. Notice what is indeed far from evident that there is a range of~$\gamma_1$ values for which~$\gamma(t)$ becomes negative at times and yet allows the system to be in a stable oscillating state. 
	
	The non-monotonic behavior of the relaxation times and the lack of simple algebraic relations to determine stability makes it hard to envisage the exotic landscape of the parameter space of the driven system containing regions that either further or forbid oscillating states. We survey and chart out some of the planes of the parameter space, for instance, ${k_1-\omega}$ plane as shown in Fig.~\hyperref[MuPlots]{2(a)}, ${\gamma_1-k_1}$ plane as shown in Fig.~\hyperref[MuPlots]{2(b)}, and ${k_1-k_0}$ plane as shown in Fig.~\hyperref[MuPlots]{2(c)}. The maps clearly indicate that as we move in these planes we could pass through a variety of terrain starting from vast stretches of stable regions to tentacles of allowed zones separated by impermissible gaps. It should be emphasized that the stable regions are identified not only by ensuring that the condition~\eqref{eff-cond-1} holds but also by confirming in parallel that the covariance matrix~\eqref{cov-mat} is positive definite.

	\subsection{Concurrence of the perturbations}
	
	Suppose we choose the driven parameters~$k, \gamma$ and~$D$ such that a normalizable asymptotic distribution in absence of the nonlinear forces exists. We still need to address whether such a choice is compatible with the condition~\eqref{cond-pert} for if it is not the case then the perturbations will destroy the oscillating state. This condition essentially ensures periodicity and boundedness of the perturbative corrections to the oscillating state. 
	
	The difficulty in determining the compatibility is that the condition~\eqref{cond-pert} involves Floquet exponents of a Hill equation and hence has to be verified explicitly. We first recast this condition to a form similar to Eq.~\eqref{eff-cond-1} that is more convenient for distinguishing various regions where the perturbative coefficients~$a_r^L$ exist and are periodic. We shall then proceed to address the issue of compatibility.
	
	Let $p_1(t)$ and $p_2(t)$ be two independent solutions of the Hill equation~\eqref{hill-eq} with the initial conditions:~$p_1(0)=1$, $p_2(0)=0$, $\dot{p}_1(0)=0$ and $\dot{p}_2(0)=1$. The corresponding monodromy matrix 
	\begin{align}\label{mon-mat-2}
		\Phi_p(T) = 
		\begin{pmatrix}
			p_1(T) & \dot{p}_1(T) \\
			p_2(T) & \dot{p}_2(T)
		\end{pmatrix}~,
	\end{align} 
	and the Floquet coefficients 
	\begin{equation}\label{floq-coeff-2}
		e^{\pm \mu_p T} = \frac{\Delta_p}{2} \pm \sqrt{ \frac{\Delta_p^2}{4}-1 }~,
	\end{equation}
	where the trace 
	\begin{equation}\label{tr-2}
		\Delta_p = p_1(T) + \dot{p}_2(T)~.
	\end{equation}
	Substituting Eq.~\eqref{floq-coeff-2} in condition~\eqref{cond-pert} leads to the relation
	\begin{equation}\label{eff-cond-2}
		|\Delta_p| < 2\cosh\left( \frac{1}{2} \overline{\gamma}_p T \right)~,
	\end{equation}
	which when satisfied guarantees the coexistence of the perturbations.

	Since the quantities~$\Delta_p, \overline{\gamma}_p$ and~$\Delta$ are fixed by the parameters~$k, \gamma$ and~$D$ it is natural to ask whether there are regions in the parameter space where the two relations~\eqref{eff-cond-1} and~\eqref{eff-cond-2} are mutually incompatible. Astonishingly we found strong numerical evidence to the contrary and noticed the following two relations,
	\begin{eqnarray}
		\label{sup-1}&\overline{\gamma}_p = - \overline{\gamma} ~,  \\
		\label{sup-2}&\Delta_p = \Delta~,
	\end{eqnarray}
	for any given~$k, \gamma$ and~$D$ that supports an unperturbed oscillating state. We have in fact sampled a class of periodic driving that included higher harmonics up to tenth order and numerically explored a 64 parameter space either randomly or by continuously varying one of the parameters. 
	Though $\gamma_p$ depends on two other additional parameters $k$ and $D$ in a nontrivial way, yet we find that its time average is heedless of their being. Though the functions~$\nu$ and $\nu_p$ are found to be wildly different from each other in general, the corresponding Floquet exponents appear to be equal without any exception. The prime reason for not foreseeing these relations a priori is that unlike~$\nu$ the function~$\nu_p$ in addition depends on $D$ not only implicitly through $\Sigma^{-1}$ but also explicitly. 
	
	We now prove that the relations~\eqref{sup-1} and~\eqref{sup-2} indeed hold. The first relation easily follows once we note that the dynamics of the determinant~$|\Sigma|$ of the asymptotic covariant matrix~$\Sigma$ which can be obtained from Eq.~\eqref{mom-2-as} can be written as
	\begin{equation}\label{det-dyn}
		\frac{d}{dt} \ln|\Sigma| = -2 \gamma + 2D (\Sigma^{-1})_{22}~.
	\end{equation}
	The right hand side of the above equation reduces to~$-(\gamma+\gamma_p)$ upon using Eq.~\eqref{gp-kp}. The time-period average of the left hand side vanishes since the moments of the asymptotic distribution are $T$-periodic, and thus follows the relation~\eqref{sup-1}.
	
	We need a chain of arguments to establish the second relation. Using Eqs.~\eqref{mom-2-as} it is straightforward to obtain the equations of motion for all the elements of the inverse covariance matrix~$S$ and write down the following system of nonlinear equations,
	\begin{equation}\label{inv-corr-eqn}
		\frac{d}{dt} {\bf c} = M({\bf c}) {\bf c} + {\bf d}({\bf c})
	\end{equation}
	where the transpose of ${\bf c}$ and ${\bf d}({\bf c})$ are
	\begin{eqnarray}\label{c-d-def}
		&{\bf c}^T = \Big(  S_{11} , 2 S_{12} , S_{22} \Big)~,\nonumber \\
		&{\bf d}({\bf c})^T = 2 D \Big(  S^2_{12} , 2 S_{12} S_{22} , S^2_{22} \Big)~,
	\end{eqnarray}
	respectively, and the matrix~$M({\bf c})$ is
	\begin{align}\label{M-def}
		M({\bf c})=
		\begin{pmatrix}
			0 & k'  & 0 \\
			-2 & \gamma' & 2 k' \\
			0 & -1 & 2 \gamma'
		\end{pmatrix}~,
	\end{align}  
	where ${k' = k -2D S_{12} }$ and ${\gamma' = \gamma - 2 D S_{22} }$. In the limit~$t \to \infty$, the matrix~$S(t)$ approaches the inverse of the asymptotic covariance matrix~$\Sigma(t)$ and hence ${k' \to k_p}$, ${\gamma' \to \gamma_p}$, the vector~${\bf d}({\bf c})$ becomes $T$-periodic and the matrix
	\begin{equation}\label{M-asy}
		M({\bf c}) \to  M_{\infty} := \left[ - \mathbf{J}^{-}_{2} +\frac{\gamma_p }{2} \left( 2  I_2 +\mathbf{J}_{2} \right) + k_p  \mathbf{J}^{+}_{2}\right]^T .
	\end{equation}
	The asymptotic solution of Eq.~\eqref{inv-corr-eqn} is unique and bounded since the asymptotic unperturbed distribution is unique and normalizable. 
	Any solution of the equation for a given initial condition approaches the asymptotic solution, while the differences between the solutions vanish asymptotically at a rate dictated by the Floquet exponents. The dynamics of the difference~$\delta {\bf c} $ of infinitesimally separated solutions follows from Eq.~\eqref{inv-corr-eqn} and is given by
	\begin{equation}\label{deviate-eq}
		\frac{d}{dt} \delta {\bf c} = M({\bf c}) \delta{\bf c}~,
	\end{equation}
	which asymptotically takes the exact form as Eq.~\eqref{a-hom} satisfied by ${\bf a}_L$ for $L=2$. Hence, using Eq.~\eqref{Fexp-L}, we conclude that the Floquet exponents associated with the elements of the inverse covariance matrix are~${\mu_2^{(r)} = \overline{\gamma}_p + (2-2r)\mu_p}$, where ${r=0,1,2}$.  The variation of the covariance matrix~$\delta S^{-1}$ is related to the variation of the inverse covariance matrix~$\delta S$ by the relation 
	\begin{equation}\label{var-C}
		{\delta S^{-1} = - S^{-1} {\delta S} S^{-1} \to - \Sigma {\delta S} \Sigma} . 
	\end{equation}
	The Floquet exponents associated with the covariance matrix $\delta S^{-1}$ can be obtained independently and as mentioned earlier are ${-{\overline{\gamma} + (2-2r)\mu}}$. Now using Eqs.~\eqref{var-C} and~\eqref{sup-1}, we conclude~$\mu_p = \pm \mu$ and thus establish Eq.~\eqref{sup-2}.
	
	Essentially, the two conditions~\eqref{cond-pert} and~\eqref{cond-mu-g0} or the corresponding equivalent conditions~\eqref{eff-cond-2} and~\eqref{eff-cond-1} are not just compatible with each other but are in fact one and the same. In other words, we find that the perturbations can coexist with the oscillating states in the entire domain of their existence.
	
	\section{Concluding comments}\label{conc}
	
	We have considered a periodically driven Brownian particle under nonlinear forces and analysed the time dependence of the asymptotic distribution. The unperturbed Brownian particle is known to exist in an oscillating state under the conditions that the inequality~\eqref{cond-mu-g0} holds and that the covariance matrix is positive definite. To any order in the nonlinear perturbations, we find that the oscillating state either can sustain or is destroyed depending on whether or not the condition~\eqref{cond-pert} holds. The reason that it is the same condition that arbitrates the existence of the oscillating state at every non-zero order of perturbation is due to the presence of the underlying $SL_2$ symmetry. We have essentially formulated the perturbative analysis suitable both for identifying the symmetry, as given in Eq.~\eqref{a-hom}, and for obtaining exactly the asymptotic distribution in terms of the solution of the Hill equation~\eqref{hill-eq}. We have obtained the formal expression of the first order coefficients~$\mathbf{a}_L(t)$ of the asymptotic distribution as given in Eqs.~\eqref{asy-sol-L}  and~\eqref{asy-sol-L-2}, and outlined the procedure to determine the higher order coefficients.
	
	We have addressed the issue of the stability of the oscillating state which is essentially related to the compatibility of the two conditions~\eqref{cond-mu-g0} and~\eqref{cond-pert}. These conditions implicitly depend on the driving parameters in a nontrivial way involving Floquet exponents of the corresponding Hill equations. We have charted out some of the terrains of the oscillating states in the parameter space of driving involving first harmonics so as to explicitly demonstrate the nontrivial relation between driving parameters and the stated conditions. More importantly, we have proved the equivalence of these conditions inspite of their implicit dependence on the parameters and established that the oscillating states are stable against nonlinear perturbations to all orders of perturbation.
	
	The ubiquity of the driven systems and their access to experiments is a strong motivation to study the effect of different perturbations on the oscillating states. The perturbative formulation developed here could prove valuable not only in understanding the properties of these states but also in establishing the appropriate description of driven physical systems.  
	
	Since a variety of macroscopic systems undergo stochastic dynamics similar to that of Brownian motion, these systems when driven has the possibility to exist in stable oscillating states. The necessary conditions that we obtained could be effective in identifying the type of driving required to maintain an oscillating state.
	
	The analysis that we have employed here can also be easily extended to study driven stochastic systems possessing other symmetries. It would be interesting to know the role of symmetry on the necessary and sufficient conditions under which the oscillating states can exist and sustain.


\begin{thebibliography}{23}%
	\makeatletter
	\providecommand \@ifxundefined [1]{%
		\@ifx{#1\undefined}
	}%
	\providecommand \@ifnum [1]{%
		\ifnum #1\expandafter \@firstoftwo
		\else \expandafter \@secondoftwo
		\fi
	}%
	\providecommand \@ifx [1]{%
		\ifx #1\expandafter \@firstoftwo
		\else \expandafter \@secondoftwo
		\fi
	}%
	\providecommand \natexlab [1]{#1}%
	\providecommand \enquote  [1]{``#1''}%
	\providecommand \bibnamefont  [1]{#1}%
	\providecommand \bibfnamefont [1]{#1}%
	\providecommand \citenamefont [1]{#1}%
	\providecommand \href@noop [0]{\@secondoftwo}%
	\providecommand \href [0]{\begingroup \@sanitize@url \@href}%
	\providecommand \@href[1]{\@@startlink{#1}\@@href}%
	\providecommand \@@href[1]{\endgroup#1\@@endlink}%
	\providecommand \@sanitize@url [0]{\catcode `\\12\catcode `\$12\catcode
		`\&12\catcode `\#12\catcode `\^12\catcode `\_12\catcode `\%12\relax}%
	\providecommand \@@startlink[1]{}%
	\providecommand \@@endlink[0]{}%
	\providecommand \url  [0]{\begingroup\@sanitize@url \@url }%
	\providecommand \@url [1]{\endgroup\@href {#1}{\urlprefix }}%
	\providecommand \urlprefix  [0]{URL }%
	\providecommand \Eprint [0]{\href }%
	\providecommand \doibase [0]{https://doi.org/}%
	\providecommand \selectlanguage [0]{\@gobble}%
	\providecommand \bibinfo  [0]{\@secondoftwo}%
	\providecommand \bibfield  [0]{\@secondoftwo}%
	\providecommand \translation [1]{[#1]}%
	\providecommand \BibitemOpen [0]{}%
	\providecommand \bibitemStop [0]{}%
	\providecommand \bibitemNoStop [0]{.\EOS\space}%
	\providecommand \EOS [0]{\spacefactor3000\relax}%
	\providecommand \BibitemShut  [1]{\csname bibitem#1\endcsname}%
	\let\auto@bib@innerbib\@empty
	\bibitem [{\citenamefont {Higashikawa}\ \emph {et~al.}(2018)\citenamefont
		{Higashikawa}, \citenamefont {Fujita},\ and\ \citenamefont
		{Sato}}]{Higashikawa2018}%
	\BibitemOpen
	\bibfield  {author} {\bibinfo {author} {\bibfnamefont {S.}~\bibnamefont
			{Higashikawa}}, \bibinfo {author} {\bibfnamefont {H.}~\bibnamefont
			{Fujita}},\ and\ \bibinfo {author} {\bibfnamefont {M.}~\bibnamefont {Sato}},\
	}\bibfield  {title} {\bibinfo {title} {Floquet engineering of classical
			systems},\ }\href {https://arxiv.org/abs/1810.01103} {\bibfield  {journal}
		{\bibinfo  {journal} {arXiv:1810.01103}\ } (\bibinfo {year}
		{2018})}\BibitemShut {NoStop}%
	\bibitem [{\citenamefont {Salerno}\ \emph {et~al.}(2016)\citenamefont
		{Salerno}, \citenamefont {Ozawa}, \citenamefont {Price},\ and\ \citenamefont
		{Carusotto}}]{Salerno2016}%
	\BibitemOpen
	\bibfield  {author} {\bibinfo {author} {\bibfnamefont {G.}~\bibnamefont
			{Salerno}}, \bibinfo {author} {\bibfnamefont {T.}~\bibnamefont {Ozawa}},
		\bibinfo {author} {\bibfnamefont {H.~M.}\ \bibnamefont {Price}},\ and\
		\bibinfo {author} {\bibfnamefont {I.}~\bibnamefont {Carusotto}},\ }\bibfield
	{title} {\bibinfo {title} {Floquet topological system based on
			frequency-modulated classical coupled harmonic oscillators},\ }\href
	{https://doi.org/10.1103/PhysRevB.93.085105} {\bibfield  {journal} {\bibinfo
			{journal} {Phys. Rev. B}\ }\textbf {\bibinfo {volume} {93}},\ \bibinfo
		{pages} {085105} (\bibinfo {year} {2016})}\BibitemShut {NoStop}%
	\bibitem [{\citenamefont {Bukov}\ \emph {et~al.}(2015)\citenamefont {Bukov},
		\citenamefont {D'Alessio},\ and\ \citenamefont {Polkovnikov}}]{Bukov2015}%
	\BibitemOpen
	\bibfield  {author} {\bibinfo {author} {\bibfnamefont {M.}~\bibnamefont
			{Bukov}}, \bibinfo {author} {\bibfnamefont {L.}~\bibnamefont {D'Alessio}},\
		and\ \bibinfo {author} {\bibfnamefont {A.}~\bibnamefont {Polkovnikov}},\
	}\bibfield  {title} {\bibinfo {title} {Universal high-frequency behavior of
			periodically driven systems: from dynamical stabilization to floquet
			engineering},\ }\href {https://doi.org/10.1080/00018732.2015.1055918}
	{\bibfield  {journal} {\bibinfo  {journal} {Advances in Physics}\ }\textbf
		{\bibinfo {volume} {64}},\ \bibinfo {pages} {139} (\bibinfo {year}
		{2015})}\BibitemShut {NoStop}%
	\bibitem [{\citenamefont {Eckardt}\ and\ \citenamefont
		{Anisimovas}(2015)}]{Eckardt2015}%
	\BibitemOpen
	\bibfield  {author} {\bibinfo {author} {\bibfnamefont {A.}~\bibnamefont
			{Eckardt}}\ and\ \bibinfo {author} {\bibfnamefont {E.}~\bibnamefont
			{Anisimovas}},\ }\bibfield  {title} {\bibinfo {title} {High-frequency
			approximation for periodically driven quantum systems from a floquet-space
			perspective},\ }\href {https://doi.org/10.1088/1367-2630/17/9/093039}
	{\bibfield  {journal} {\bibinfo  {journal} {New Journal of Physics}\ }\textbf
		{\bibinfo {volume} {17}},\ \bibinfo {pages} {093039} (\bibinfo {year}
		{2015})}\BibitemShut {NoStop}%
	\bibitem [{\citenamefont {Kohler}\ \emph {et~al.}(1997)\citenamefont {Kohler},
		\citenamefont {Dittrich},\ and\ \citenamefont {H\"anggi}}]{Kohler1997}%
	\BibitemOpen
	\bibfield  {author} {\bibinfo {author} {\bibfnamefont {S.}~\bibnamefont
			{Kohler}}, \bibinfo {author} {\bibfnamefont {T.}~\bibnamefont {Dittrich}},\
		and\ \bibinfo {author} {\bibfnamefont {P.}~\bibnamefont {H\"anggi}},\
	}\bibfield  {title} {\bibinfo {title} {Floquet-markovian description of the
			parametrically driven, dissipative harmonic quantum oscillator},\ }\href
	{https://doi.org/10.1103/PhysRevE.55.300} {\bibfield  {journal} {\bibinfo
			{journal} {Phys. Rev. E}\ }\textbf {\bibinfo {volume} {55}},\ \bibinfo
		{pages} {300} (\bibinfo {year} {1997})}\BibitemShut {NoStop}%
	\bibitem [{\citenamefont {{Lewis}}\ and\ \citenamefont
		{{Riesenfeld}}(1969)}]{Lewis1969}%
	\BibitemOpen
	\bibfield  {author} {\bibinfo {author} {\bibfnamefont {J.}~\bibnamefont
			{{Lewis}}, \bibfnamefont {H.~R.}}\ and\ \bibinfo {author} {\bibfnamefont
			{W.~B.}\ \bibnamefont {{Riesenfeld}}},\ }\bibfield  {title} {\bibinfo {title}
		{{An Exact Quantum Theory of the Time-Dependent Harmonic Oscillator and of a
				Charged Particle in a Time-Dependent Electromagnetic Field}},\ }\href
	{https://doi.org/10.1063/1.1664991} {\bibfield  {journal} {\bibinfo
			{journal} {Journal of Mathematical Physics}\ }\textbf {\bibinfo {volume}
			{10}},\ \bibinfo {pages} {1458} (\bibinfo {year} {1969})}\BibitemShut
	{NoStop}%
	\bibitem [{\citenamefont {Brandner}\ and\ \citenamefont
		{Seifert}(2016)}]{Brandner2016}%
	\BibitemOpen
	\bibfield  {author} {\bibinfo {author} {\bibfnamefont {K.}~\bibnamefont
			{Brandner}}\ and\ \bibinfo {author} {\bibfnamefont {U.}~\bibnamefont
			{Seifert}},\ }\bibfield  {title} {\bibinfo {title} {Periodic thermodynamics
			of open quantum systems},\ }\href
	{https://doi.org/10.1103/PhysRevE.93.062134} {\bibfield  {journal} {\bibinfo
			{journal} {Phys. Rev. E}\ }\textbf {\bibinfo {volume} {93}},\ \bibinfo
		{pages} {062134} (\bibinfo {year} {2016})}\BibitemShut {NoStop}%
	\bibitem [{\citenamefont {Jung}(1993)}]{Jung1993}%
	\BibitemOpen
	\bibfield  {author} {\bibinfo {author} {\bibfnamefont {P.}~\bibnamefont
			{Jung}},\ }\bibfield  {title} {\bibinfo {title} {Periodically driven
			stochastic systems},\ }\href {https://doi.org/10.1016/0370-1573(93)90022-6}
	{\bibfield  {journal} {\bibinfo  {journal} {Physics Reports}\ }\textbf
		{\bibinfo {volume} {234}},\ \bibinfo {pages} {175} (\bibinfo {year}
		{1993})}\BibitemShut {NoStop}%
	\bibitem [{\citenamefont {Brandner}\ \emph {et~al.}(2015)\citenamefont
		{Brandner}, \citenamefont {Saito},\ and\ \citenamefont
		{Seifert}}]{Brandner2015}%
	\BibitemOpen
	\bibfield  {author} {\bibinfo {author} {\bibfnamefont {K.}~\bibnamefont
			{Brandner}}, \bibinfo {author} {\bibfnamefont {K.}~\bibnamefont {Saito}},\
		and\ \bibinfo {author} {\bibfnamefont {U.}~\bibnamefont {Seifert}},\
	}\bibfield  {title} {\bibinfo {title} {Thermodynamics of micro- and
			nano-systems driven by periodic temperature variations},\ }\href
	{https://doi.org/10.1103/PhysRevX.5.031019} {\bibfield  {journal} {\bibinfo
			{journal} {Phys. Rev. X}\ }\textbf {\bibinfo {volume} {5}},\ \bibinfo {pages}
		{031019} (\bibinfo {year} {2015})}\BibitemShut {NoStop}%
	\bibitem [{\citenamefont {Dutta}\ and\ \citenamefont
		{Barma}(2003)}]{Dutta2003}%
	\BibitemOpen
	\bibfield  {author} {\bibinfo {author} {\bibfnamefont {S.~B.}\ \bibnamefont
			{Dutta}}\ and\ \bibinfo {author} {\bibfnamefont {M.}~\bibnamefont {Barma}},\
	}\bibfield  {title} {\bibinfo {title} {Asymptotic distributions of
			periodically driven stochastic systems},\ }\href
	{https://doi.org/10.1103/PhysRevE.67.061111} {\bibfield  {journal} {\bibinfo
			{journal} {Phys. Rev. E}\ }\textbf {\bibinfo {volume} {67}},\ \bibinfo
		{pages} {061111} (\bibinfo {year} {2003})}\BibitemShut {NoStop}%
	\bibitem [{\citenamefont {Dutta}(2004)}]{Dutta2004}%
	\BibitemOpen
	\bibfield  {author} {\bibinfo {author} {\bibfnamefont {S.~B.}\ \bibnamefont
			{Dutta}},\ }\bibfield  {title} {\bibinfo {title} {Phase transitions in
			periodically driven macroscopic systems},\ }\href
	{https://doi.org/10.1103/PhysRevE.69.066115} {\bibfield  {journal} {\bibinfo
			{journal} {Phys. Rev. E}\ }\textbf {\bibinfo {volume} {69}},\ \bibinfo
		{pages} {066115} (\bibinfo {year} {2004})}\BibitemShut {NoStop}%
	\bibitem [{\citenamefont {Wang}\ and\ \citenamefont
		{Sch{\"{u}}tte}(2015)}]{Wang2015}%
	\BibitemOpen
	\bibfield  {author} {\bibinfo {author} {\bibfnamefont {H.}~\bibnamefont
			{Wang}}\ and\ \bibinfo {author} {\bibfnamefont {C.}~\bibnamefont
			{Sch{\"{u}}tte}},\ }\bibfield  {title} {\bibinfo {title} {{Building markov
				state models for periodically driven non-equilibrium systems}},\ }\href
	{https://doi.org/10.1021/ct500997y} {\bibfield  {journal} {\bibinfo
			{journal} {Journal of Chemical Theory and Computation}\ }\textbf {\bibinfo
			{volume} {11}},\ \bibinfo {pages} {1819} (\bibinfo {year}
		{2015})}\BibitemShut {NoStop}%
	\bibitem [{\citenamefont {Knoch}\ and\ \citenamefont
		{Speck}(2019)}]{Knoch2019}%
	\BibitemOpen
	\bibfield  {author} {\bibinfo {author} {\bibfnamefont {F.}~\bibnamefont
			{Knoch}}\ and\ \bibinfo {author} {\bibfnamefont {T.}~\bibnamefont {Speck}},\
	}\bibfield  {title} {\bibinfo {title} {Non-equilibrium markov state modeling
			of periodically driven biomolecules},\ }\href
	{https://doi.org/10.1063/1.5055818} {\bibfield  {journal} {\bibinfo
			{journal} {The Journal of Chemical Physics}\ }\textbf {\bibinfo {volume}
			{150}},\ \bibinfo {pages} {054103} (\bibinfo {year} {2019})}\BibitemShut
	{NoStop}%
	\bibitem [{\citenamefont {Gammaitoni}\ \emph {et~al.}(1998)\citenamefont
		{Gammaitoni}, \citenamefont {H\"anggi}, \citenamefont {Jung},\ and\
		\citenamefont {Marchesoni}}]{Gammaitoni1998}%
	\BibitemOpen
	\bibfield  {author} {\bibinfo {author} {\bibfnamefont {L.}~\bibnamefont
			{Gammaitoni}}, \bibinfo {author} {\bibfnamefont {P.}~\bibnamefont
			{H\"anggi}}, \bibinfo {author} {\bibfnamefont {P.}~\bibnamefont {Jung}},\
		and\ \bibinfo {author} {\bibfnamefont {F.}~\bibnamefont {Marchesoni}},\
	}\bibfield  {title} {\bibinfo {title} {Stochastic resonance},\ }\href
	{https://doi.org/10.1103/RevModPhys.70.223} {\bibfield  {journal} {\bibinfo
			{journal} {Rev. Mod. Phys.}\ }\textbf {\bibinfo {volume} {70}},\ \bibinfo
		{pages} {223} (\bibinfo {year} {1998})}\BibitemShut {NoStop}%
	\bibitem [{\citenamefont {Kim}\ \emph {et~al.}(2010)\citenamefont {Kim},
		\citenamefont {Talkner}, \citenamefont {Lee},\ and\ \citenamefont
		{H\"anggi}}]{Kim2010}%
	\BibitemOpen
	\bibfield  {author} {\bibinfo {author} {\bibfnamefont {C.}~\bibnamefont
			{Kim}}, \bibinfo {author} {\bibfnamefont {P.}~\bibnamefont {Talkner}},
		\bibinfo {author} {\bibfnamefont {E.~K.}\ \bibnamefont {Lee}},\ and\ \bibinfo
		{author} {\bibfnamefont {P.}~\bibnamefont {H\"anggi}},\ }\bibfield  {title}
	{\bibinfo {title} {Rate description of fokker-planck processes with
			time-periodic parameters},\ }\href
	{https://doi.org/https://doi.org/10.1016/j.chemphys.2009.10.027} {\bibfield
		{journal} {\bibinfo  {journal} {Chemical Physics}\ }\textbf {\bibinfo
			{volume} {370}},\ \bibinfo {pages} {277} (\bibinfo {year}
		{2010})}\BibitemShut {NoStop}%
	\bibitem [{\citenamefont {Fiore}\ and\ \citenamefont
		{de~Oliveira}(2019)}]{Fiore2019}%
	\BibitemOpen
	\bibfield  {author} {\bibinfo {author} {\bibfnamefont {C.~E.}\ \bibnamefont
			{Fiore}}\ and\ \bibinfo {author} {\bibfnamefont {M.~J.}\ \bibnamefont
			{de~Oliveira}},\ }\bibfield  {title} {\bibinfo {title} {Entropy production
			and heat capacity of systems under time-dependent oscillating temperature},\
	}\href {https://doi.org/10.1103/PhysRevE.99.052131} {\bibfield  {journal}
		{\bibinfo  {journal} {Phys. Rev. E}\ }\textbf {\bibinfo {volume} {99}},\
		\bibinfo {pages} {052131} (\bibinfo {year} {2019})}\BibitemShut {NoStop}%
	\bibitem [{\citenamefont {Koyuk}\ \emph {et~al.}(2018)\citenamefont {Koyuk},
		\citenamefont {Seifert},\ and\ \citenamefont {Pietzonka}}]{Koyuk2018}%
	\BibitemOpen
	\bibfield  {author} {\bibinfo {author} {\bibfnamefont {T.}~\bibnamefont
			{Koyuk}}, \bibinfo {author} {\bibfnamefont {U.}~\bibnamefont {Seifert}},\
		and\ \bibinfo {author} {\bibfnamefont {P.}~\bibnamefont {Pietzonka}},\
	}\bibfield  {title} {\bibinfo {title} {A generalization of the thermodynamic
			uncertainty relation to periodically driven systems},\ }\href
	{https://doi.org/10.1088/1751-8121/aaeec4} {\bibfield  {journal} {\bibinfo
			{journal} {Journal of Physics A: Mathematical and Theoretical}\ }\textbf
		{\bibinfo {volume} {52}},\ \bibinfo {pages} {02LT02} (\bibinfo {year}
		{2018})}\BibitemShut {NoStop}%
	\bibitem [{\citenamefont {Oberreiter}\ \emph {et~al.}(2019)\citenamefont
		{Oberreiter}, \citenamefont {Seifert},\ and\ \citenamefont
		{Barato}}]{Oberreiter2019}%
	\BibitemOpen
	\bibfield  {author} {\bibinfo {author} {\bibfnamefont {L.}~\bibnamefont
			{Oberreiter}}, \bibinfo {author} {\bibfnamefont {U.}~\bibnamefont
			{Seifert}},\ and\ \bibinfo {author} {\bibfnamefont {A.~C.}\ \bibnamefont
			{Barato}},\ }\bibfield  {title} {\bibinfo {title} {Subharmonic oscillations
			in stochastic systems under periodic driving},\ }\href
	{https://doi.org/10.1103/PhysRevE.100.012135} {\bibfield  {journal} {\bibinfo
			{journal} {Phys. Rev. E}\ }\textbf {\bibinfo {volume} {100}},\ \bibinfo
		{pages} {012135} (\bibinfo {year} {2019})}\BibitemShut {NoStop}%
	\bibitem [{\citenamefont {Tociu}\ \emph {et~al.}(2019)\citenamefont {Tociu},
		\citenamefont {Fodor}, \citenamefont {Nemoto},\ and\ \citenamefont
		{Vaikuntanathan}}]{Tociu2019}%
	\BibitemOpen
	\bibfield  {author} {\bibinfo {author} {\bibfnamefont {L.}~\bibnamefont
			{Tociu}}, \bibinfo {author} {\bibfnamefont {E.}~\bibnamefont {Fodor}},
		\bibinfo {author} {\bibfnamefont {T.}~\bibnamefont {Nemoto}},\ and\ \bibinfo
		{author} {\bibfnamefont {S.}~\bibnamefont {Vaikuntanathan}},\ }\bibfield
	{title} {\bibinfo {title} {How dissipation constrains fluctuations in
			nonequilibrium liquids: Diffusion, structure, and biased interactions},\
	}\href {https://doi.org/10.1103/PhysRevX.9.041026} {\bibfield  {journal}
		{\bibinfo  {journal} {Phys. Rev. X}\ }\textbf {\bibinfo {volume} {9}},\
		\bibinfo {pages} {041026} (\bibinfo {year} {2019})}\BibitemShut {NoStop}%
	\bibitem [{\citenamefont {Awasthi}\ and\ \citenamefont
		{Dutta}(2020)}]{Awasthi2020}%
	\BibitemOpen
	\bibfield  {author} {\bibinfo {author} {\bibfnamefont {S.}~\bibnamefont
			{Awasthi}}\ and\ \bibinfo {author} {\bibfnamefont {S.~B.}\ \bibnamefont
			{Dutta}},\ }\bibfield  {title} {\bibinfo {title} {Periodically driven
			harmonic langevin systems},\ }\href
	{https://doi.org/10.1103/PhysRevE.101.042106} {\bibfield  {journal} {\bibinfo
			{journal} {Phys. Rev. E}\ }\textbf {\bibinfo {volume} {101}},\ \bibinfo
		{pages} {042106} (\bibinfo {year} {2020})}\BibitemShut {NoStop}%
	\bibitem [{\citenamefont {Fulton}\ and\ \citenamefont
		{Harris}(2004)}]{Fulton2004}%
	\BibitemOpen
	\bibfield  {author} {\bibinfo {author} {\bibfnamefont {W.}~\bibnamefont
			{Fulton}}\ and\ \bibinfo {author} {\bibfnamefont {J.}~\bibnamefont
			{Harris}},\ }\href {https://doi.org/10.1007/978-1-4612-0979-9} {\emph
		{\bibinfo {title} {Representation theory: A First Course}}},\ Vol.\ \bibinfo
	{volume} {129}\ (\bibinfo  {publisher} {Springer-Verlag New York},\ \bibinfo
	{year} {2004})\BibitemShut {NoStop}%
	\bibitem [{\citenamefont {Magnus}\ and\ \citenamefont
		{Winkler}(2013)}]{Magnus2013}%
	\BibitemOpen
	\bibfield  {author} {\bibinfo {author} {\bibfnamefont {W.}~\bibnamefont
			{Magnus}}\ and\ \bibinfo {author} {\bibfnamefont {S.}~\bibnamefont
			{Winkler}},\ }\href@noop {} {\emph {\bibinfo {title} {Hill's equation}}}\
	(\bibinfo  {publisher} {Courier Corporation},\ \bibinfo {year}
	{2013})\BibitemShut {NoStop}%
	\bibitem [{\citenamefont {Eastham}(1975)}]{Eastham1975}%
	\BibitemOpen
	\bibfield  {author} {\bibinfo {author} {\bibfnamefont {M.}~\bibnamefont
			{Eastham}},\ }\href@noop {} {\emph {\bibinfo {title} {Spectral theory and
				differential equations}}}\ (\bibinfo  {publisher} {Springer},\ \bibinfo
	{year} {1975})\BibitemShut {NoStop}%
\end{thebibliography}
\end{document}